\documentstyle[prb,aps,multicol,eqsecnum,epsf]{revtex}
\begin{document}
\title{Energy Spectrum of Neutral Collective Excitations 
in Striped Hall States} 
\author{T. Aoyama, K. Ishikawa, Y. Ishizuka, and N. Maeda}
\address{
Department of Physics, Hokkaido University, 
Sapporo 060-0810, Japan}
\date{\today}
\maketitle
\begin{abstract}
We investigate neutral collective excitations in the striped Hall state. 
In the striped Hall state, the magnetic translation and rotation symmetries 
are spontaneously broken. 
Using the commutation relation between charges and currents corresponding 
to the broken and unbroken symmetry, 
the existence of the gapless neutral excitation is proved. 
The spectrum of the neutral collective excitation at the half-filled 
third Landau level is obtained in the single mode approximation. 
We find the periodic line nodes in the spectrum. 
The spectrum is compared with the particle-hole excitation spectrum 
in the Hartree-Fock approximation. 
\end{abstract}
\draft

\begin{multicols}{2}

\section{Introduction}

The two-dimensional electron system under a strong magnetic field 
shows astonishing diversity in experiments and theories. 
The quantized Hall conductance\cite{Kli,Tsui} is observed in 
the incompressible quantum liquid state.\cite{Laugh}
Fractional quantum Hall states (FQHS) have anyonic quasi-particles 
in the excited states.\cite{Arov} 
Around the half-filled lowest Landau level, the composite Fermi 
liquid state\cite{Comp} which has an isotropic Fermi surface is observed. 
At the half-filled second Landau level, on the other hand, 
the fractionally quantized Hall conductance is observed;\cite{Pan} 
it is suggested that the state has an energy gap due to the Cooper 
pairing of the electrons.\cite{Grei,Maeda1} 

Recently, highly anisotropic states, which have extremely anisotropic 
longitudinal resistances, have been observed around the half-filled third 
and higher Landau levels.\cite{Lilly,Du} 
It is believed that the anisotropic state is the striped Hall 
state which is a unidirectional charge density wave in the mean field 
theory.\cite{Kou,Moe} 
The anisotropy of the resistance is naturally explained by the 
anisotropic Fermi surface 
in the magnetic Brillouin zone.\cite{Imo,Maeda2} 
It was predicted that fluctuation effects turn the striped Hall state 
into the smectic or nematic liquid 
crystal.\cite{Fra,Mac,Cote,Fog,Wex,Rad} 
A low energy theory was studied in the coupled Luttinger liquid 
picture using the Hartree-Fock approximation.\cite{Anna} 
A peculiar response to external modulations was predicted 
in a perturbation theory\cite{Ao} and numerical calculations.\cite{Yos} 

In the highly anisotropic states, underlying symmetries are spontaneously 
broken and Nambu-Goldstone (NG) modes should appear. 
It is important to find the property of the NG mode. 
In this paper, we investigate the property of the NG mode 
in striped Hall states. 
In the two-dimensional system under a uniform magnetic field, 
the system has the magnetic translation and rotation symmetry. 
In the striped Hall state, a magnetic translation in one direction is 
broken to the discrete translation and the rotation is also 
broken to the $\pi$-rotation, spontaneously. 
We prove the Goldstone theorem for the striped Hall state using 
the commutation relation between charges and currents corresponding to the 
broken and unbroken symmetry. 
The theorem shows that a gapless neutral excitation exists 
and couples with the density operator. 
The spectrum of the neutral collective excitation is obtained in 
the single mode approximation numerically at the half-filled 
third Landau level. 
The spectrum has a multiple line node structure and cusps. 
Furthermore, the spectrum has anisotropic feature, that is, 
in one direction it resembles the liquid Helium 
spectrum with the phonon and roton minimum, and in another 
direction it resembles the FQHS spectrum. 
To show the validity of the single mode approximation in the present 
system, we compare the spectrum with the particle-hole 
excitation spectrum in the Hartree-Fock approximation. 

The paper is organized as follows. 
Conserved currents in the two-dimensional system under a 
uniform magnetic field are clarified in Sec.~II. 
In Sec.~III, a mean field theory for the striped Hall state 
is presented. 
Goldstone theorem for the striped Hall state is proved in Sec.~IV. 
The spectrum for the neutral collective excitation in the 
striped Hall state is obtained in the single mode approximation 
and compared with the particle-hole excitation spectrum in the 
Hartree-Fock approximation in Sec.~V. 
Summary is given in Sec.~VI. 

\section{Conserved currents in a uniform magnetic field}

Let us consider the two-dimensional electron system in a uniform magnetic 
field $B=\partial_x A_y-\partial_y A_x$. 
We ignore the spin degree of freedom and use the natural unit 
($\hbar=c=1$) in this paper. 
We introduce two sets of coordinates, relative coordinates and 
guiding center coordinates. 
The relative coordinates are defined by 
\begin{equation}
\xi={1\over eB}(-i\partial_y+eA_y),\ 
\eta=-{1\over eB}(-i\partial_x+eA_x).
\end{equation}
The guiding center coordinates are defined by 
\begin{equation}
X=x-\xi,\ Y=y-\eta.
\end{equation}
These coordinates satisfy the following commutation 
relations, 
\begin{eqnarray}
&[X,Y]=-[\xi,\eta]=i/eB,\nonumber\\
&[X,\xi]=[X,\eta]=[Y,\xi]=[Y,\eta]=0.
\end{eqnarray}
The operators $X$ and $Y$ are the generators of the magnetic 
translations of the one-particle state in $-y$ direction and 
$x$ direction respectively. 
The angular momentum $J$ is written as 
\begin{equation}
J={eB\over2}(\xi^2+\eta^2-X^2-Y^2).
\end{equation}
$J$ is the generator of the rotation of the one-particle state. 

The total Hamiltonian $H$ for the interacting charged particles 
is the sum of the free Hamiltonian $H_0$ and the Coulomb interaction 
Hamiltonian $H_{\rm int}$ as follows,
\begin{eqnarray}
H&=&H_0+H_{\rm int},\nonumber\\
H_0&=&\int d{\bf r} \Psi^\dagger({\bf r}){m \omega_c^2\over2}
(\xi^2+\eta^2)\Psi^\dagger({\bf r}),\\
H_{\rm int}&=&{1\over2}\int d{\bf r}d{\bf r}'\Psi^\dagger({\bf r})
\Psi^\dagger({\bf r}')V({\bf r}-{\bf r}')\Psi({\bf r}')\Psi({\bf r}),
\nonumber
\end{eqnarray}
where $\Psi$ is the electron field operator, $\omega_c=eB/m$ and 
$V({\bf r})=q^2/r$ 
($q^2=e^2/4 \pi \epsilon $, $\epsilon$ is the 
dielectric constant) for the Coulomb potential. 

Conserved charges are obtained as the spatial integral of the zeroth 
component of the Noether currents for the symmetries. 
We define the conserved charges, $Q$, $Q_X$, $Q_Y$, and $Q_J$ for 
U(1), magnetic translations, and rotation symmetry respectively, 
as follows, 
\begin{eqnarray}
&Q=\int j^0({\bf r})d{\bf r},&\nonumber\\
&Q_X=\int j_X^0({\bf r})d{\bf r},\ 
Q_Y=\int j_Y^0({\bf r})d{\bf r},&\\
&Q_J=\int j_J^0({\bf r})d{\bf r}.&\nonumber
\end{eqnarray}
Noether currents are defined by 
\begin{eqnarray}
j^\mu&=&{\rm Re}(\Psi^\dagger v^\mu\Psi),\nonumber\\
j_X^\mu&=&{\rm Re}(\Psi^\dagger v^\mu X \Psi)-{1\over eB}\delta_y^\mu
{\cal L},\\
j_Y^\mu&=&{\rm Re}(\Psi^\dagger v^\mu Y \Psi)+{1\over eB}\delta_x^\mu 
{\cal L},\nonumber\\
j_J^\mu&=&{\rm Re}(\Psi^\dagger v^\mu J \Psi)+
\epsilon_{0\mu i}x^i{\cal L},\nonumber
\end{eqnarray}
where $v^\mu=(1,{\bf v})$, ${\bf v}=\omega_c(-\eta,\xi)$, 
Re means real part, and $\cal L$ is the 
Lagrangian density for the total Hamiltonian $H$. 
These charges commute with the total Hamiltonian $H$ and 
obey the following commutation relations, 
\begin{eqnarray}
&[Q_X,Q_Y]&={i\over eB}Q,\nonumber\\ 
&[Q_J,Q_X]&=iQ_Y,\\ 
&[Q_J,Q_Y]&=-iQ_X.\nonumber
\end{eqnarray}
The U(1) charge $Q$ commutes with all conserved charges. 
We assume that $Q$ is not broken and the ground state is the eigenstate 
of $Q$ as $Q\vert 0\rangle=N_e\vert 0\rangle$ where $N_e$ is a number of 
electrons. 
It should be noted that these symmetries are not broken 
in the Laughlin state.\cite{Laugh} 
Hence the NG mode does not exist. 
The Laughlin state is the eigenstate of $Q_J$ and 
annihilated by $Q_X+iQ_Y$. 

The commutation relations between the conserved charges and 
the current density operators read,
\begin{eqnarray}
&[Q_X,j^\mu]&=-{i\over eB}\partial_y j^\mu,\ 
[Q_Y,j^\mu]={i\over eB}\partial_x j^\mu,\label{com}\\ 
&[Q_X,j_X^\mu]&=-{i\over eB}\partial_y j_X^\mu,\ 
[Q_Y,j_X^\mu]={i\over eB}\partial_x j_X^\mu-{i\over eB}j^\mu,
\nonumber\\ 
&[Q_X,j_Y^\mu]&=-{i\over eB}\partial_y j_Y^\mu+{i\over eB}j^\mu,\ 
[Q_Y,j_Y^\mu]={i\over eB}\partial_x j_Y^\mu.\nonumber
\end{eqnarray}
The U(1) charge $Q$ commutes with all Noether currents. 
If the expectation value of the right hand side of these equation 
is not zero, corresponding symmetry is spontaneously broken. 
In the striped Hall state discussed in the next section, 
$\partial_x\langle j^\mu\rangle$ 
is not zero and the magnetic translation symmetry in $x$ direction 
is spontaneously broken. 

\section{Symmetry breaking in the striped Hall state}

In this section, we study the self-consistent solution for the striped 
Hall state in the Hartree-Fock approximation. 
The state breaks the translation and rotational symmetry 
spontaneously. 

We use the von Neumann lattice (vNL) base for the one-particle 
states.\cite{Ima} 
A discrete set of coherent states of guiding center coordinates, 
\begin{eqnarray}
&(X+iY)\vert\alpha_{mn}\rangle=z_{mn}\vert\alpha_{mn}\rangle,\\
&z_{mn}=a(mr_s+i{n\over r_s}),\ m,\ n;{\rm\ integers},\nonumber
\end{eqnarray}
is a complete set of the $(X,Y)$ space. 
These coherent states are localized at the position $a(mr_s,n/r_s)$, 
where a positive real number $r_s$ is the asymmetric parameter 
of the unit cell. 
By Fourier transforming these states, we obtain the orthonormal 
basis in the momentum representation
\begin{eqnarray}
&\vert\beta_{\bf p}\rangle=\sum_{mn} e^{ip_x m+ip_y n}
\vert\alpha_{mn}\rangle/\beta({\bf p}),\\
&\beta({\bf p})=(2{\rm Im}\tau)^{1/4}e^{{i\tau p_y^2\over4\pi}}
\vartheta_1({p_x+\tau p_y\over 2\pi}\vert\tau),\nonumber
\end{eqnarray}
where $\vartheta_1$ is a Jacobi's theta function and $\tau=ir_s^2$. 
The two-dimensional lattice momentum $\bf p$ is defined in the 
Brillouin zone (BZ) $\vert p_i\vert<\pi$. 
Next we introduce a complete set in $(\xi,\eta)$ space, that is 
the eigenstate of the one-particle free Hamiltonian, 
\begin{equation}
{m\omega^2_c\over2}(\xi^2+\eta^2)\vert f_l\rangle=
\omega_c(l+{1\over2})\vert f_l\rangle. 
\end{equation}
The Hilbert space is spanned by the direct product of these eigenstates 
\begin{equation}
\vert l,{\bf p}\rangle=\vert f_l\rangle\otimes\vert\beta_{\bf p}\rangle,
\end{equation}
where $l$ is the Landau level index and $l=0$, 1, 2, $\cdots$. 
We set $a=1$ $(eB=2\pi)$ in the following calculation for simplicity. 

Electron field operator is expanded by the vNL base as
\begin{equation}
\Psi({\bf r})=\sum_{l=0}^{\infty}\int_{\rm BZ}{d^2p\over(2\pi)^2}
b_l({\bf p})\langle {\bf r}\vert l,{\bf p}\rangle,
\end{equation}
where $b_l$ is the anti-commuting annihilation operator. 
$b_l({\bf p})$ obeys the non-trivial boundary condition, 
$b_l({\bf p}+2\pi{\bf N})=e^{i\phi(p,N)}b_l({\bf p})$, 
where $\phi(p,N)=\pi(N_x+N_y)-N_y p_x$ and 
${\bf N}=(N_x,N_y)$ are intergers. 
The Fourier transform of the density $\rho({\bf k})=
\int d{\bf r}e^{i{\bf k}\cdot{\bf r}}j^0({\bf r})$ is written as
\begin{eqnarray}
\rho({\bf k})&=&\sum_{ll'}
\int_{\rm BZ}{d^2p\over(2\pi)^2}b_l^\dagger({\bf p})
b_{l'}({\bf p}-\hat{\bf k})\langle f_l\vert e^{ik_x\xi+ik_y\eta}\vert 
f_{l'}\rangle\nonumber\\
&&\times e^{-{i\over4\pi}{\hat k}_x(2p_y-{\hat k}_y)},
\end{eqnarray}
where $\hat{\bf k}=(r_s k_x,k_y/r_s)$. 
Conserved charges are written in the momentum representation as
\begin{eqnarray}
Q&=&\sum_l\int_{\rm BZ}{d^2p\over(2\pi)^2}b_l^\dagger({\bf p})
b_l({\bf p}),\nonumber\\
Q_X&=&r_s\sum_l\int_{\rm BZ}{d^2p\over(2\pi)^2}b_l^\dagger({\bf p})
(i{\partial\over\partial p_x}-{p_y\over2\pi})
b_l({\bf p}),\nonumber\\
Q_Y&=&{1\over r_s}\sum_l\int_{\rm BZ}{d^2p\over(2\pi)^2}b_l^\dagger({\bf p})
i{\partial\over\partial p_y}b_l({\bf p}),
\label{mom}\\
Q_J&=&\sum_l\int_{\rm BZ}{d^2p\over(2\pi)^2}b_l^\dagger({\bf p})
[l+{1\over2}\nonumber\\
&&+\pi\{r_s^2 (i{\partial\over\partial p_x}-{p_y\over2\pi})^2+
r_s^{-2}(i{\partial\over\partial p_y})^2\}]b_l({\bf p}).\nonumber
\end{eqnarray}
As seen in Eq.~(\ref{mom}), 
the magnetic translation in the real space is equivalent to 
the magnetic translation in the momentum space. 
The free kinetic energy is quenched in a magnetic field and 
the one-particle spectrum becomes flat. 
Therefore the free system in a magnetic field is 
translationally invariant in the momentum space. 
We show that there exists the Fermi surface in the mean field 
solution for the striped Hall state. 
The existence of the Fermi surface indicates the violation 
of the translational symmetry in the momentum space or 
in the real space. 

The mean field state is constructed as
\begin{equation}
\vert0\rangle=N_1 \prod_{{\bf p}\in {\rm F.S.}}
b_l^\dagger({\bf p})\vert{\rm vac}\rangle,
\label{mfs}
\end{equation}
where F.S. means Fermi sea, $N_1$ is a normalization constant, and 
$\vert{\rm vac}\rangle$ is the vacuum state in which the $l-1$ th and 
lower Landau levels are fully occupied. 
The mean field of the two-point function for the electron field 
is given by
\begin{eqnarray}
\langle0\vert b^\dagger_l({\bf p})b_{l'}({\bf p}')
\vert0\rangle&=&\delta_{ll'}\theta(\mu-\epsilon_{\rm HF}({\bf p}))\\
&&\times
\sum_N(2\pi)^2\delta({\bf p}-{\bf p}'+2\pi {\bf N})e^{i\phi(p,N)},
\nonumber
\end{eqnarray}
where $\mu$ is a chemical potential and one-particle energy 
$\epsilon_{\rm HF}({\bf p})$ is determined self-consistently in the 
Hartree-Fock approximation.\cite{Imo,Maeda2} 
We assume that the magnetic field $B$ is so strong that 
Landau level mixing effects can be neglected.\cite{YM}  
Then we use the Hamiltonian projected to the $l$ th Landau level 
\begin{eqnarray}
P_l H P_l&=&\omega_c N_e^* (l+{1\over2})+ H^{(l)},\\
H^{(l)}&=&P_l H_{\rm int}P_l,\nonumber
\nonumber
\end{eqnarray}
where $P_l$ is the projection operator to the $l$ th Landau level and 
$N_e^*$ is a number of electrons occupying the $l$ th Landau level. 
The free kinetic term is quenched and the total Hamiltonian is reduced 
to the Coulomb interaction term projected to the $l$ th Landau level, 
$H^{(l)}$. 

It was shown that the mean field state with Fermi sea $\vert p_y\vert<\pi 
\nu_*$ satisfies the self-consistency equation at the filling factor 
$\nu=l+\nu_*$ ($0<\nu_*<1$) in the Hartree-Fock approximation
\cite{Imo,Maeda2} 
and corresponds to the striped Hall 
state whose density is uniform in $y$ direction and periodic 
with a period $r_s$ in $x$ direction.\cite{Maeda2} 
The one-particle energy $\epsilon_{\rm HF}({\bf p})$ 
depends only on $p_y$ in this self-consistent solution. 
The Fermi sea and corresponding charge density distribution in $x$-$y$ 
space are sketched in Fig.~1. 
The period of stripe is $r_s$. 
The electric current flows along each stripes. 
Note that the true charge density and current density 
distribution in the mean field theory 
are not sharp as shown in the figure but fuzzy.\cite{Maeda2} 
As seen in Fig.~1 (b), $\partial_x\langle j^0\rangle$ and $\partial_x 
\langle j^y\rangle$ are not zero and the magnetic translation 
in $x$-direction and rotation symmetry are spontaneously broken, 
but a finite magnetic translation in $x$-direction by the period 
of stripe $r_s$ and $\pi$ rotation symmetry are preserved. 

\begin{figure}
\epsfxsize=3in\centerline{\epsffile{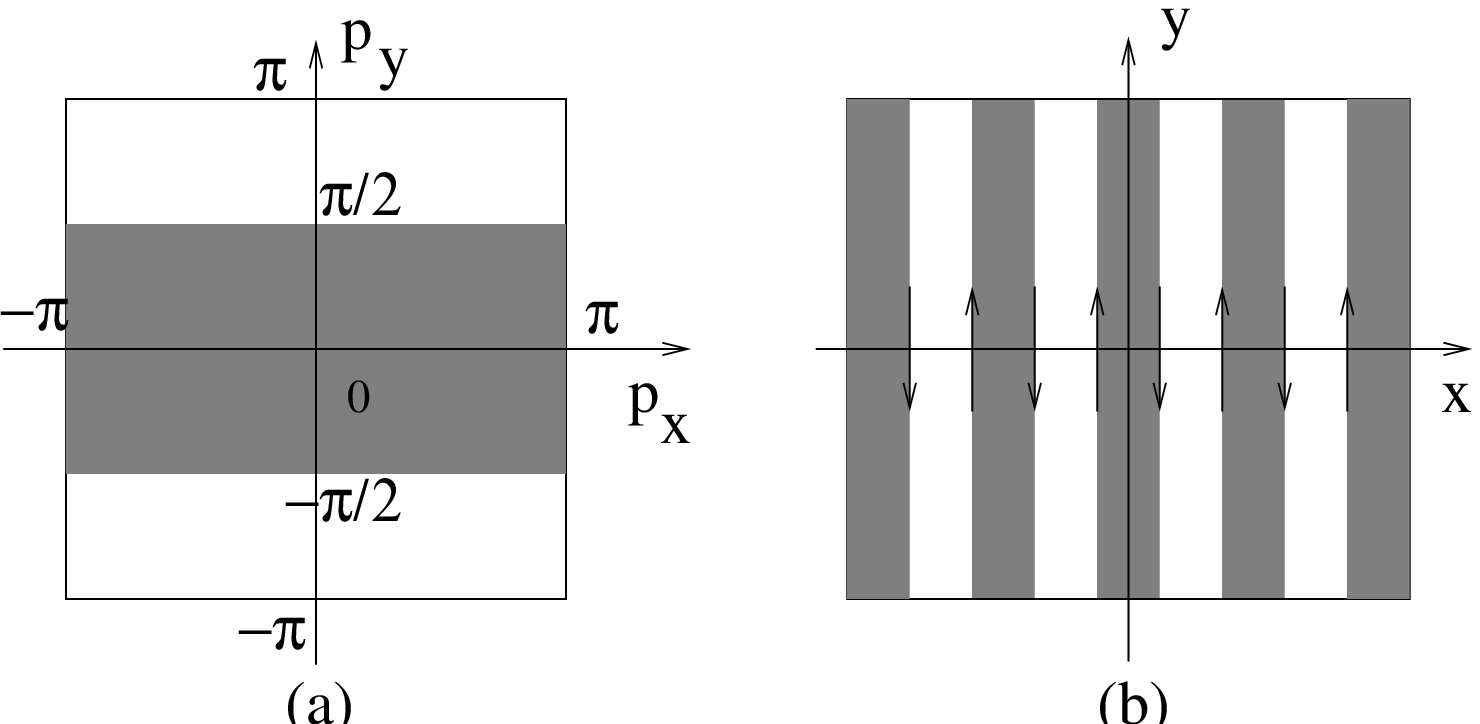}}
Fig.~1. (a) Fermi sea for the striped Hall state at the half-filling. 
(b) Schematical charge density distribution in $x$-$y$ space. 
The arrows indicate the direction of electric current. 
\end{figure}

Hartree-Fock energy per area for the striped Hall state is calculated as
\begin{eqnarray}
E_{\rm HF}&=&\langle0\vert H^{(l)}\vert0\rangle/L^2\\
&=&{1\over2}\int_{-\pi/2}^{\pi/2}{dp_y\over2\pi}\epsilon_{\rm HF}(p_y),
\nonumber
\end{eqnarray}
where $L^2$ is an area of the system. 
$E_{\rm HF}$ depends on the period of the stripe $r_s$ and the 
optimal value of $r_s$ is determined by minimizing $E_{\rm HF}$. 
At the filling factor $\nu=2+{1\over2}$, 
the optimal value is $r_s^*=2.474$.\cite{Maeda2} 

\section{Goldstone theorem for the striped Hall state}

In this section, we assume that the magnetic translation in $x$ direction 
and rotation symmetries are broken, and the magnetic translation in $y$ 
direction, finite translation in $x$ direction with a period $r_s$ and 
$\pi$ rotation are preserved in the striped Hall state. 
The Goldstone theorem for the striped Hall state is proven under 
these general assumptions. 

For definite discussions, the periodic boundary condition in $(X,Y)$ 
space is imposed for $\langle\alpha_{mn}\vert\beta_{\bf p}\rangle$ and 
${\bf p}$ is discretized as ${\bf p}=({2\pi n_x/L_x},{2\pi n_y/L_y})$, where 
$n_x$, $n_y$ are integers and $L_x$, $L_y$ are numbers of lattice sites 
in the $x$ direction and $y$ direction, respectively. 
Therefore the well-defined translation operators in the Hilbert space 
are $e^{i2\pi n_x{Q_X/r_s L_x}}$ and 
$e^{i2\pi n_y{r_s Q_Y/ L_y}}$. 
For the striped Hall state, unbroken operators are $Q$, 
$e^{i2\pi {Q_X/r_s L_x}}$, $e^{i2\pi r_s Q_Y}$ and 
$e^{i\pi Q_J}$. 
Then we can specify the state using the eigenvalues of the unbroken 
operators as
\begin{eqnarray}
H\vert n\rangle&=&E_n\vert n\rangle,\nonumber\\
e^{i2\pi{Q_X\over r_s L_x}}\vert n\rangle&=&
e^{i2\pi{Q_X^{(n)}\over r_s L_x}}\vert n\rangle,
\label{ei}\\
e^{i2\pi r_s Q_Y}\vert n\rangle&=&e^{i2\pi r_s Q_Y^{(n)}}\vert
n\rangle,\nonumber\\
Q\vert n\rangle&=&N_e\vert n\rangle,\nonumber
\end{eqnarray}
where $n\geq0$ and $\vert 0\rangle$ is the ground 
state for the striped Hall state. 
For the ground state $\vert 0\rangle$, we can show that 
$e^{i2\pi{Q_X^{(0)}/r_s L_x}}=\pm1$ and
$e^{i2\pi r_s Q_Y^{(0)}}=\pm1$. 
We can also show that $e^{i\pi Q_J}\vert 0\rangle=\pm\vert 0\rangle$ 
(See Appendix A). 
The thermodynamic limit $L_x$, $L_y\rightarrow\infty$ should be 
taken at the last stage of calculations. 
For example, $Q_X$ is defined by $\lim_{L_x\rightarrow\infty}
{r_s L_x\over2\pi i}(e^{i2\pi {Q_X/r_s L_x}}-1)$. 
Since $Q_X$ and $Q_Y$ correspond to the magnetic translation in 
$-y$ and $x$ direction respectively, the eigenvalues $Q_X^{(n)}$ 
and $Q_Y^{(n)}$ are regarded as a kind of momentum in $-y$ and $x$ 
direction respectively. 

Let us show the Goldstone theorem for the striped Hall state. 
An expectation value for $\vert0\rangle$ of the commutation relations 
Eq.~(\ref{com}) for the broken charge $Q_Y$ reads
\begin{eqnarray}
{i\over2\pi}\partial_x\langle j^{0}({\bf r},t)\rangle&=&
\langle0\vert [Q_Y,j^{0}({\bf r},t)]\vert0\rangle
\label{Gold}\\
&=&\int d{\bf r}' \langle0\vert [j_Y^0({\bf r}',t'),j^{0}({\bf r},t)]
\vert0\rangle\nonumber\\
&=&\sum_n\int d{\bf r}'\{\langle0\vert j_Y^0({\bf r}',t')\vert n\rangle
\langle n\vert j^{0}({\bf r},t)\vert0\rangle\nonumber\\
&&\qquad\qquad -h.c.\},\nonumber
\end{eqnarray}
where $h.c.$ means hermitian conjugate. 
The left-hand side of this equation is not zero and does not depend on 
time $t$. 
The right-hand side of this equation is rewritten by using translation 
operators in $t$ and $\bf r$. 
Note that $j_Y^0$ is transformed covariantly under the magnetic 
translation in $y$ direction, that is, 
\begin{eqnarray}
e^{i2\pi\Delta y Q_X}j_Y^0({\bf r})e^{-i2\pi\Delta y Q_X}&=&
j^0_Y({\bf r}+(0,\Delta y))\label{cov}\\
&&-\Delta y j^0({\bf r}+(0,\Delta y)).\nonumber
\end{eqnarray}
In the $x$ direction, we use 
the notation ${\bf r}=(N_x r_s+{\bar x},y)$, where $N_x$ 
is an integer and $0<{\bar x}<r_s$. 
Using the relation $[Q,j^0]=0$ 
and Eqs.~(\ref{ei}) and (\ref{cov}), 
the right-hand side of Eq.~(\ref{Gold}) is rewritten as
\begin{eqnarray}
&&\sum_{n>0}\int_0^{r_s}{d{\bar x}'\over r_s}
[\{\langle0\vert j_Y^0(({\bar x}',0),0)
\vert n\rangle\langle n\vert j^{0}(({\bar x},0),0)\vert0\rangle
\label{Go}\\
&&\times e^{i(t'-t)(E_n-E_0)}-h.c.\}
\delta(Q^{(0)}_X-Q_X^{(n)})\delta_{1\over r_s}(Q_Y^{(0)}-Q_Y^{(n)})
\nonumber\\
&&+{1\over2\pi i}\{\langle0\vert j^0(({\bar x}',0),0)
\vert n\rangle\langle n\vert j^{0}(({\bar x},0),0)\vert0\rangle
\nonumber\\
&&\times e^{i(t'-t)(E_n-E_0)}+h.c.\}
\delta'(Q^{(0)}_X-Q_X^{(n)})\delta_{1\over r_s}(Q_Y^{(0)}-Q_Y^{(n)})]
\nonumber
\end{eqnarray}
where $\delta_{1\over r_s}(Q^{(0)}_Y-Q_Y^{(n)})$ is a periodic Dirac's 
delta-function with a period $1/r_s$. 
The derivative of the Dirac's delta-function appears in the second term. 
We assume the completeness $1=\sum_n\vert n\rangle\langle n\vert
=\vert 0\rangle\langle 0\vert+\sum_\alpha 
\int d{\bf q}\vert {\bf q},\alpha\rangle\langle {\bf q},\alpha\vert$. 
Here $\vert {\bf q},\alpha\rangle$ is an excited state with 
eigenvalues ${\bf Q}^{(n)}={\bf Q}^{(0)}+{\bf q}$, 
$E_n=E_0+\Delta E_\alpha({\bf q})$, and species index $\alpha$. 
Then Eq.~(\ref{Go}) is rewritten as
\begin{eqnarray}
&&\sum_\alpha\int_0^{r_s}{d{\bar x}'\over r_s}
[\{\langle0\vert j_Y^0(({\bar x}',0),0)
\vert{\bf 0},\alpha\rangle\langle{\bf 0},\alpha\vert j^{0}(({\bar x},0),0)
\vert0\rangle\nonumber\\
&&\times e^{i(t'-t)\Delta E_\alpha({\bf 0})}-h.c.\}
\label{Gol}\\
&&+\{{i\over2\pi}{\partial\over\partial q_x}
\langle0\vert j^0(({\bar x}',0),0)
\vert{\bf q},\alpha\rangle\langle{\bf 0},\alpha\vert j^{0}(({\bar x},0),0)
\vert0\rangle\nonumber\\
&&\times e^{i(t'-t)\Delta E_\alpha({\bf 0})}-h.c.\}_{{\bf q}=0}],
\nonumber
\end{eqnarray}
where we use the relation $\langle 0\vert Q\vert{\bf q},\alpha\rangle=0$. 
The left-hand side of Eq.~(\ref{Gold}) does not depend on $t$ and 
equals Eq.~(\ref{Gol}). 
Hence there exists the NG mode 
whose excitation energy $\Delta E_{\rm NG}({\bf q})$ goes to zero as 
${\bf q}\rightarrow0$. 
Since the matrix elements in Eq.~(\ref{Gol}) must not be zero, 
NG mode $\vert {\bf q},{\rm NG}\rangle$ belongs to the neutral charged 
sector and couples with the striped Hall state through the density operator, 
that is,
\begin{equation}
\langle {\bf q},{\rm NG}\vert j^0\vert0\rangle\neq0,
\end{equation}
in ${\bf q}\rightarrow0$. 

This NG mode is a phonon due to breaking of a magnetic translation. 
NG modes (phonon)\cite{And} also appear in the Wigner crystal 
in the absence of a magnetic field. 
In this case, the second term in Eq.~(\ref{Gol}) is absent. 
It would be interesting if this term is observed in the quantum Hall 
system. 
In the next section, we investigate the excitation spectrum using 
the density operator $j^0$ which couples with the NG mode. 

\section{Neutral collective excitations in the striped Hall state}

In this section we calculate the spectrum for a neutral collective 
excitation at the half-filled third Landau level using the single 
mode approximation. 
This method was used for the Liquid Helium by Feynman first
\cite{Fey} and applied to the FQHS later.\cite{Gir} 
The single mode approximation is successful in the FQHS because 
the backflow problem is absent for the electron states projected 
to the $l$ th Landau level. 

First we define the Fourier transformed density operator 
in guiding center coordinates as
\begin{eqnarray}
\rho_*({\bf k})&=&P_l\int d{\bf r}\Psi^\dagger({\bf r})
e^{ik_x X+ik_y Y}\Psi({\bf r})P_l\\
&=&\int_{\rm BZ}{d^2p\over(2\pi)^2}b_l^\dagger({\bf p})
b_{l}({\bf p}-\hat{\bf k})e^{-{i\over4\pi}{\hat k}_x(2p_y-{\hat k}_y)}.
\nonumber
\end{eqnarray}
Density operator $\rho({\bf k})$ is related to $\rho_*({\bf k})$ as 
$P_l \rho({\bf k})P_l=e^{-k^2/8\pi}L_l(k^2/4\pi)\rho_*({\bf k})$, 
where $L_l$ is the Laguerre polynomial. 
Using $\rho_*({\bf k})$, $H^{(l)}$ is written as
\begin{equation}
H^{(l)}={1\over2}\int{d^2k\over(2\pi)^2}\rho_*({\bf k})
v_l(k)\rho_*(-{\bf k}),
\end{equation}
where $v_l(k)=e^{-k^2/4\pi}(L_l(k^2/4\pi))^2 2\pi q^2/k$ for 
the Coulomb potential. 
It is well-known that the density operators projected to the 
Landau level are non-commutative, that is, 
\begin{equation}
[\rho_*({\bf k}),\rho_*({\bf k}')]=-2i\sin\left(
{{\bf k}\times{\bf k}'\over4\pi}\right)
\rho_*({\bf k}+{\bf k}').
\label{rho}
\end{equation}
Using this relation, we obtain the following commutation relations, 
\begin{eqnarray}
&[Q_X,\rho_*({\bf k})]&=-{k_y\over2\pi}\rho_*({\bf k}),\\
&[Q_Y,\rho_*({\bf k})]&={k_x\over2\pi}\rho_*({\bf k}).
\nonumber
\end{eqnarray}

Therefore the state defined by
\begin{equation}
\vert{\bf k}\rangle=\rho_*({\bf k})\vert 0\rangle
\label{exc}
\end{equation}
is an eigenstate of $e^{i2\pi{Q_X/r_s L_x}}$ and 
$e^{i2\pi r_s Q_Y}$ with eigenvalues $e^{i2\pi(Q_X^{(0)}-{k_y\over2\pi})
/r_s L_x}$ and $e^{i2\pi r_s(Q_Y^{(0)}+{k_x\over2\pi})}$ 
respectively. 
The quantum number $\bf q$ for the excited state is related to $\bf k$ 
as $k_i=2\pi\epsilon^{ij}q_j$. 
We use the state $\vert{\bf k}\rangle$ as a neutral collective 
excitation state in the single mode approximation. 

The variational excitation energy $\Delta({\bf k})$ is written as
\begin{eqnarray}
\Delta({\bf k})&=&{\langle{\bf k}\vert (H^{(l)}-E_0)\vert{\bf k}\rangle
\over\langle{\bf k}\vert{\bf k}\rangle}=
{f({\bf k})\over s({\bf k})},\nonumber\\
f({\bf k})&=&\langle0\vert [\rho_*(-{\bf k}),[H^{(l)},\rho_*({\bf k})]]
\vert0\rangle/2N_e^*,\\
s({\bf k})&=&\langle0\vert\rho_*(-{\bf k})\rho_*({\bf k})
\vert0\rangle/N_e^*.\nonumber
\end{eqnarray}
$s({\bf k})$ is the so-called static structure factor. 
To derive these expressions, we use the relation $f(-{\bf k})=f({\bf k})$ 
and $s(-{\bf k})=s({\bf k})$ due to $\pi$ rotation symmetry. 
See appendix A. 
Using the commutation relation (\ref{rho}), $f({\bf k})$ is written as 
\begin{eqnarray}
f({\bf k})&=&2\int{d^2k'\over(2\pi)^2}v_l(k')\sin^2
\left({{\bf k}'\times{\bf k}\over4\pi}\right)
\label{fk}\\
&&\times\{s({\bf k}+{\bf k}')-s({\bf k}')\}\nonumber
\end{eqnarray}
From Eq.~(\ref{fk}) 
the variational excitation energy is calculable if we know 
the static structure factor $s({\bf k})$. 
Using the mean field state (\ref{mfs}) for the striped Hall state, 
$s({\bf k})$ becomes
\begin{eqnarray}
s({\bf k})&=&{1\over\nu_*}\int_{\rm BZ}{d^2 p\over(2\pi)^2}
\theta(\mu-\epsilon_{\rm HF}(p_y))
\label{ssf}\\
&&\times
\{1-\theta(\mu-\epsilon_{\rm HF}(p_y-{\hat k}_y))\}
\nonumber\\
&&+\sum_{N_x}{(2\pi)^2\over\nu_*}\delta({\hat k}_x+2\pi N_x)
\delta({\hat k}_y)
\left({\sin(\pi\nu_* N_x)\over \pi N_x}\right)^2.
\nonumber
\end{eqnarray}
The first term behaves as $\vert k_y\vert/2\pi r_s\nu_*$ at small 
$k_y$ and periodic in $k_y$ direction with a period $2 \pi r_s$. 
The second term has sharp peaks at ${\bf k}=(2\pi N_x/r_s,0)$. 
These peaks correspond to the period for the stripe of the charge density. 
This behavior in the static structure factor was also obtained in the 
numerical calculation in a small system.\cite{Rez,sheng} 
The numerical results for the energy spectrum $\Delta$ for $\nu=2+1/2$ 
are shown in Figs.~2-4. 

\begin{figure}
\epsfxsize=3in\centerline{\epsffile{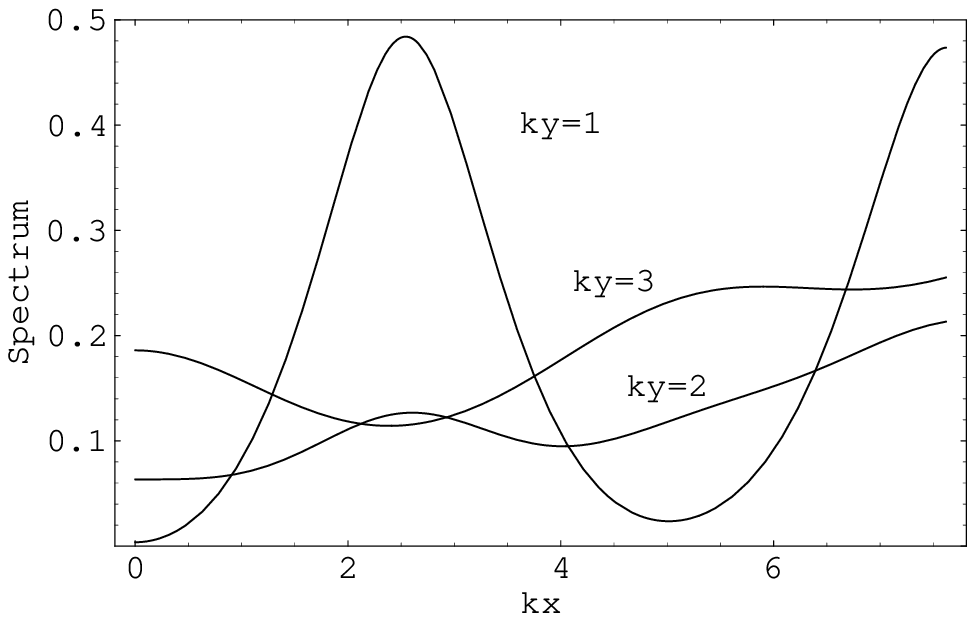}}
Fig.~2. Energy spectrum $\Delta$ at $0<k_x<6\pi/r_s$ 
for $k_y=1$, 2, 3, $\nu=2+1/2$ in the single 
mode approximation. Note that the spectrum remains finite at 
$k_x=0$ for $k_y=1$. 
The unit of $\bf k$ is $a^{-1}$ and the unit of spectrum is 
$q^2/a$. 
The same units are used in Figs.~3-5. 
\end{figure}
\begin{figure}
\epsfxsize=3in\centerline{\epsffile{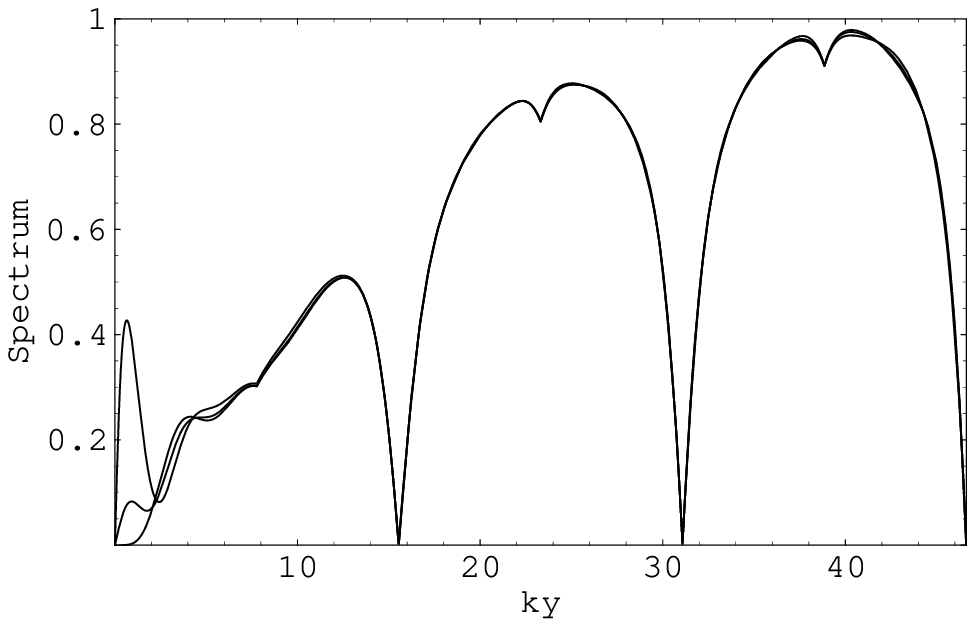}}
Fig.~3. Energy spectrum $\Delta$ at $0<k_y<6\pi r_s$ 
for $k_x=0$, 1, 2, $\nu=2+1/2$ in the single 
mode approximation. The energy gap vanishes at $k_y=2\pi N_y r_s$ and 
cusps appear at $k_y=\pi (2 N_y+1) r_s$. 
\end{figure}
\begin{figure}
\epsfxsize=3in\centerline{\epsffile{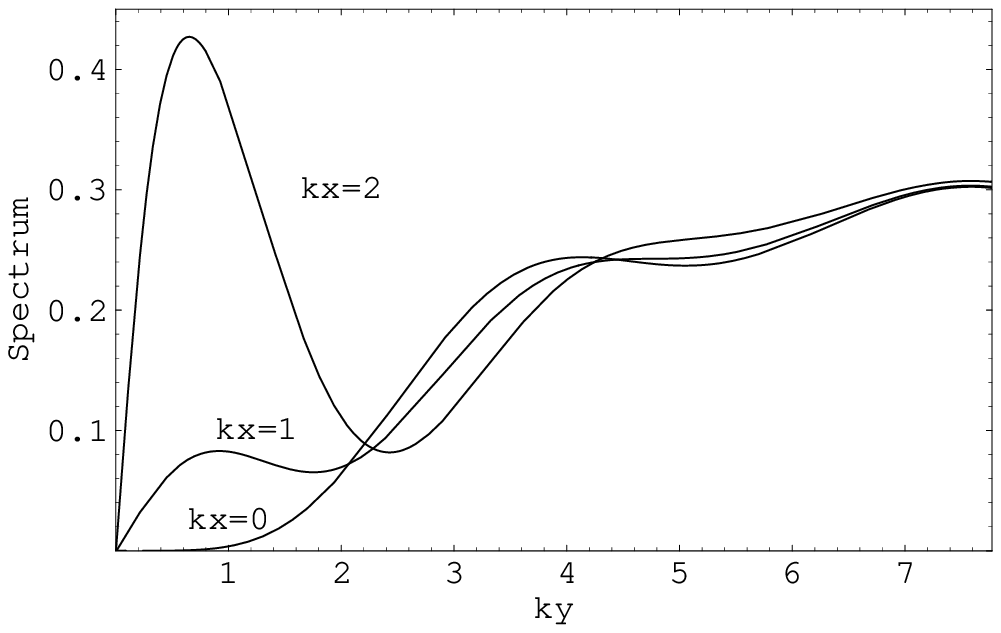}}
Fig.~4. Enlarged picture of the energy spectrum $\Delta$ at 
$0<k_y<\pi r_s$ for $k_x=0$, 1, 2, $\nu=2+1/2$ 
in the single mode approximation. 
\end{figure}

As predicted by the Goldstone theorem in the previous section, 
the energy spectrum obtained by the single mode approximation 
is gapless at ${\bf k}=0$. 
In fact, we can show that $\Delta({\bf k})$ behaves as 
\begin{equation}
\Delta({\bf k})=\vert k_y\vert(A k_x^2+B k_y^4+{\rm O}(k_x^2 k_y^2,k_x^4)),
\label{Del}
\end{equation}
at small $\bf k$ for a short-range potential. 
The reason for the absence of $\vert k_y\vert^3$ term is that the 
coefficient of $\vert k_y\vert^3$ vanishes by the condition 
${\partial E_{\rm HF}\over\partial r_s}=0$ at $r_s=r_s^*$. 
For the Coulomb potential case, the logarithmic corrections are 
added in Eq.~(\ref{Del}).  
See appendix B. 

In addition to the gapless structure of the energy spectrum 
at ${\bf k}=0$ due to 
NG mode, the results obtained by the single mode approximation 
present more rich structure. 
First we notice a periodic line node structure, that is, 
the energy spectrum vanishes at $k_y=2\pi N_y r_s$. 
Except for $k_x=0$, the dispersion at line node is linear. 
Second we notice that the energy spectrum has cusps at 
$k_y=\pi(2 N_y+1)r_s$. 
This is caused by the presence of the Fermi surface in the 
mean field state (\ref{mfs}), which produces cusps in the first term of 
Eq.~(\ref{ssf}). 
Third the $k_x$ dependence of the spectrum becomes weak at the 
large wave number. 
Fourth the spectrum has distinct features in different directions. 
Fig.~2 shows that the spectrum in $k_x$ direction has a similarity with 
the collective excitation in FQHS\cite{Gir} which has an energy gap at 
any wave number and magneto-roton structure. 
Fig.~4 shows that the spectrum in $k_y$ direction has a similarity with 
the collective excitation in Liquid Helium which has phonon and roton 
spectrum. 
Fourth feature comes from the anisotropy of the Fermi sea. 
As seen in Fig.~1 (a), in $p_x$ direction, the electron state is fully 
filled and has an energy gap. 
Therefore the system resembles FQHS in this direction. 
In $p_y$ direction, on the other hand, there is a Fermi surface and 
electron state is gapless. 
First three features are understandable from comparison with 
the particle-hole excitation spectrum in the Hartree-Fock approximation. 

For the large wave number $\vert k\vert >2\pi/a$, it is expected that 
the excitation state becomes close to the free particle state. 
In the FQHS, the neutral excitation spectrum becomes close to the 
pair creation energy for the quasi-particle and quasi-hole.\cite{Lau2} 
In the striped Hall state, the one-particle energy is $\epsilon_{\rm HF}
(p_y)$ in the Hartree-Fock approximation and 
the free part of Hamiltonian becomes
\begin{equation}
H_{\rm HF}=\int_{\rm BZ}{d^2p\over(2\pi)^2}\epsilon_{\rm HF}(p_y)
b_l^\dagger({\bf p})b_l({\bf p}).
\end{equation}
Therefore the excitation spectrum $\Delta_{\rm ph}({\bf k})$ for the 
state of particle-hole pair $b_l^\dagger({\bf p}+\hat{\bf k})
b_l({\bf p})\vert0\rangle$ is bounded as
\begin{equation}
\Delta_{\rm ph}({\bf k})\geq\vert\epsilon_{\rm HF}(\pi\nu_*+{\hat k}_y)-
\epsilon_{\rm HF}(\pi\nu_*)\vert.
\end{equation}
$\pi\nu_*$ is a Fermi wave number. 
Lower bound of the particle-hole excitation spectrum for 
$\nu=2+1/2$ is shown in Fig.~5. 
\begin{figure}
\epsfxsize=3in\centerline{\epsffile{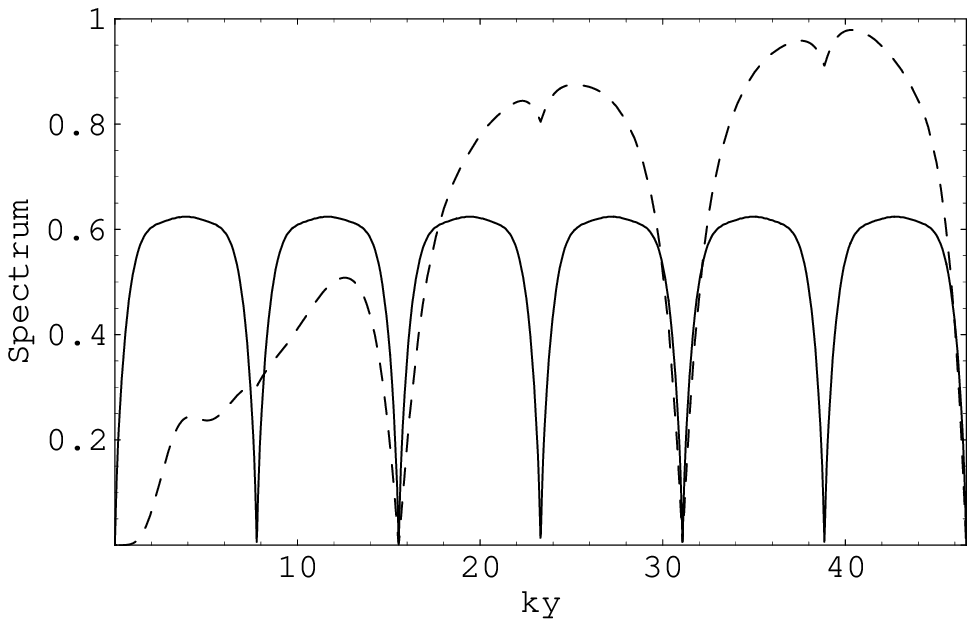}}
Fig.~5. Lower bound of the particle-hole excitation spectrum 
$\Delta_{\rm ph}$ for $0<k_y<6 \pi r_s$ (solid line). 
The spectrum vanishes at $k_y=\pi N_y r_s$. 
For comparison, the energy spectrum $\Delta$ for $k_x=0$, $\nu=2+1/2$ 
in the single mode approximation is also shown (dashed line). 
\end{figure}

As seen in this figure the particle-hole excitation spectrum vanishes 
and its slope diverges\cite{Imo} at $k_y=\pi N_y r_s$. 
See Appendix B. 
This singularity is caused by the divergence of Fermi velocity 
$\partial \epsilon_{\rm HF}/\partial p_y$ at Fermi momentum. 
The spectrum $\Delta$ is smaller than $\Delta_{\rm ph}$ and 
close to the $\Delta_{\rm ph}$ near $k_y=2\pi N_y r_s$ for a large 
wave number. 
The existence of the Fermi surface leads to zero of $\Delta_{\rm ph}$ 
at $k_y=\pi (2N_y+1) r_s$, because the excitation energy for a hole inside 
the Fermi sea near the one Fermi surface and a particle outside the Fermi 
sea near the other Fermi surface is close to zero. 
The spectrum $\Delta$ is larger than $\Delta_{\rm ph}$ near 
$k_y=\pi (2N_y+1) r_s$ and has cusps. 
Hence the low energy excitation is well-approximated by the $\Delta$ 
near $k_y=2\pi N_y r_s$ and by $\Delta_{\rm ph}$ near 
$k_y=\pi (2N_y+1) r_s$. 
In a usual Fermi system, the single mode approximation seems not to work 
well because of the particle-hole excitation near the Fermi surface 
at low energies. 
In the present case, however, the low energy excitation near the 
Fermi surface is suppressed by the divergence of Fermi velocity 
at Fermi surface. 
Hence the single mode approximation is expected to work well in the 
present system. 

At a finite $k_x$, the spectrum is linear in $k_y$ at small $\bf k$, 
which is similar to the one obtained by the hydrodynamics of 
quantum Hall smectics.\cite{Fog} 
At $k_x=0$, however, the spectrum is proportional to $k_y^5$, 
which disagrees with 
the results in the hydrodynamics of quantum Hall smectics. 
This discrepancy may be resolved by higher order correction to the 
static structure factor or by modifying the effective Hamiltonian 
in the hydrodynamics. 
At large $\bf k$, in which the hydrodynamic theory cannot be applied, 
the periodic line node structure is found in the single mode approximation 
for the first time. 

\section{Summary}

We have studied neutral collective excitations in the striped Hall state. 
The Goldstone theorem has been proved in a system with the 
magnetic translation symmetry. 
We have obtained the neutral excitation spectrum 
in the striped Hall state using the single mode approximation. 
The spectrum has a new rich structure. 
The neutral collective mode includes the NG mode due to 
spontaneous breaking of magnetic translation symmetry. 
The spectrum is highly anisotropic, that is, the dispersion in $k_x$ 
direction is similar to that of FQHS and the dispersion in $k_y$ is 
similar to that of Liquid Helium. 
The spectrum is compared with the particle-hole excitation spectrum. 
The former has line nodes with a linear dispersion and 
the latter has line nodes with a diverging Fermi velocity. 
We hope these excitations will be observed in experiments for the 
evidence of the striped Hall state.

\acknowledgements
Two of the authors (T. A. and N. M.) thank A. Dorsey and W. Pan
for useful discussions. 
This work was partially supported by the special Grant-in-Aid for 
Promotion of Education and Science in Hokkaido University provided by the 
Ministry of Education, Science, Sport, and Culture, Japan, and by 
the Grant-in-Aid for 
Scientific Research on Priority area of Research (B) 
(Dynamics of Superstrings and Field Theories, Grant No. 13135201) from 
the Ministry of Education, Science, Sport, and Culture, Japan. 

\appendix
\section{}
In this appendix, we prove the relation $e^{i\pi Q_J}\vert0\rangle=
\pm\vert0\rangle$, 
$f(-{\bf k})=f({\bf k})$ and $s(-{\bf k})=s({\bf k})$ for 
the striped Hall state. 
As seen in Eq.~(\ref{mom}), the generator for rotation is 
a free kinetic energy in a uniform magnetic field in the 
momentum space. 
Therefore it is convenient to expand the operator $b_l({\bf p})$ 
by the eigenstate of the free kinetic energy as
\begin{eqnarray}
b_l({\bf p})&=&\sum_{l'=0}^\infty b_{ll'}\psi_{l'}({\bf p}),\\
{\bf D}^2\psi_{l'}({\bf p})
&=&{1\over\pi}(l'+{1\over2})\psi_{l'}({\bf p}),
\nonumber
\end{eqnarray}
where ${\bf D}=(r_s (i{\partial\over\partial p_x}-{p_y\over2\pi}),
r_s^{-1}i{\partial\over\partial p_y})$, $\psi_{l'}({\bf p})$ 
is given by
\begin{eqnarray}
\psi_{l'}({\bf p})&=&N_{l'} \sum_n H_{l'} ({r_s(p_y+2\pi(n+{1\over2}))
\over\sqrt{2\pi}})\\
&&\times e^{-{r_s^2\over4\pi}(p_y+2\pi(n+{1\over2}))^2}
e^{i(n+{1\over2})p_x+i\pi n},\nonumber
\end{eqnarray}
$N_{l'}$ is a normalization constant, $H_l$ is the Hermite polynomial, 
and $\psi_{l'}$ has a 
symmetry $\psi_{l'}(-{\bf p})=(-1)^{l'+1}\psi_{l'}({\bf p})$. 
Using this new base, $Q_J$ reads
\begin{equation}
Q_J=\sum_{ll'}b^\dagger_{ll'}(l+l'+1)b_{ll'}.
\end{equation}
Then we obtain the following relation, 
\begin{eqnarray}
e^{i\pi Q_J}b_l({\bf p})e^{-i\pi Q_J}&=&
\sum_{l'}b_{ll'}e^{i\pi(l+l'+1)}\psi_{l'}({\bf p})
\label{pib}\\
&=&(-1)^l b_l(-{\bf p}).\nonumber
\end{eqnarray}
Since the Fermi sea for the striped Hall state is invariant under 
the $\pi$ rotation, the ground state $\vert0\rangle$ is an eigenstate of 
$e^{i\pi Q_J}$ with an eigenvalue $1$ or $-1$. 
Using Eq.~(\ref{pib}), one can prove that
\begin{eqnarray}
e^{i\pi Q_J}\rho_*({\bf k})e^{-i\pi Q_J}&=&\rho_*(-{\bf k}),
\label{pir}\\
e^{i\pi Q_J}H^{(l)}e^{-i\pi Q_J}&=&H^{(l)}.
\nonumber
\end{eqnarray}
Inserting $e^{i\pi Q_J}$ and $e^{-i\pi Q_J}$ in the definition of 
$f({\bf k})$ and $s({\bf k})$, 
one can prove that  $f(-{\bf k})=f({\bf k})$ and $s(-{\bf k})=s({\bf k})$ 
using Eq.~(\ref{pir}). 

\section{}
In this appendix, we show that the anisotropic behavior of $\Delta({\bf k})$ 
in Eq.~(\ref{Del}) comes from the condition 
${\partial E_{\rm HF}\over\partial r_s}=0$. 

Using the Fourier series expansion, the one-particle energy 
$\epsilon_{\rm HF}(p_y)$ and the Hartree-Fock energy are written as 
\begin{eqnarray}
\epsilon_{\rm HF}(p_y)&=&\sum_n v_{\rm HF}
({2\pi n\over r_s}){e^{inp_y}\sin{{\pi\over2}n}\over\pi n},
\label{hfe}\\
E_{\rm HF}&=&{1\over2}\sum_n v_{\rm HF}
({2\pi n\over r_s})\left({\sin{{\pi\over2}n}\over\pi n}\right)^2,\\
v_{\rm HF}(k)&=&v_l(k)-
\int{d^2r}v_l(2\pi r)e^{i{\bf k}\times{\bf r}}.
\nonumber
\end{eqnarray}
Furthermore, using Eq.~(\ref{ssf}), 
$f({\bf k})$ in Eq.~(\ref{fk}) is written as
\begin{eqnarray}
f({\bf k})&=&2\sum_{n={\rm odd}}\{v_{\rm HF}(\sqrt{({2\pi n\over 
r_s}+k_x)^2+k_y^2)}-v_{\rm HF}({2\pi n\over r_s})\}
\nonumber\\
&&\times\left({\sin{nk_y\over 2r_s}\over\pi n}\right)^2.
\label{f}
\end{eqnarray}
At $k_x=0$ and small $k_y$, 
it is easy to see that the coefficient of $k_y^4$ is proportional to 
${\partial E_{\rm HF}\over\partial r_s}$. 
Therefore ${\partial E_{\rm HF}\over\partial r_s}=0$ leads to 
vanishing of the coefficient of $k_y^4$ at $k_x=0$ and the lowest order 
is $k_y^6$ at $k_x=0$. 
Since $s({\bf k})$ is proportional to $\vert k_y\vert$ at small $k_y$, 
$\Delta({\bf k})={f({\bf k})\over s({\bf k})}$ 
behaves as $\vert k_y\vert^5$ at $k_x=0$ and small $k_y$. 

Note that $v_{\rm HF}({\bf k})$ behaves as $-1/k$ at 
large $k$ for the Coulomb potential. 
Hence the slope of $\epsilon_{\rm HF}(p_y)$ at $p_y=\pm\pi/2$ diverges 
logarithmically in Eq.~(\ref{hfe}). 
In Eq.~(\ref{f}), the Fourier transformation of ${\rm O}(k_x^2,k_y^2)n^{-5}$ 
with respect to $k_y$ 
results in a term of ${\rm O}(k_x^2,k_y^2)k_y^4\log\vert k_y\vert$. 
Therefore the correction of ${\rm O}(k_x^2,k_y^2)\vert k_y\vert^3
\log\vert k_y\vert$ is 
added in Eq.~(\ref{Del}) for the Coulomb potential.

\end{multicols}
\end{document}